\documentstyle[aps,prb,epsfig,manuscript]{revtex}
\begin{document}
\title{Exact ground-state for the periodic Anderson model in $D=2$ dimensions
at finite value of the interaction and absence of the direct hopping in the 
correlated f-band.}
\author{Zsolt~Gul\'acsi}
\address{ 
Department of Theoretical Physics, University of Debrecen, Poroszlay ut 6/C,
H-4010 Debrecen, Hungary }
\date{Dec., 2002}
\maketitle
\begin{abstract}
We report for the first time exact ground-states deduced for the $D=2$
dimensional generic periodic Anderson model at finite $U$, 
the Hamiltonian of the model not containing direct hopping terms 
for $f$-electrons $(t^f = 0)$. The deduced itinerant phase presents non-Fermi 
liquid properties in the normal phase, emerges for real hybridization matrix 
elements, and not requires anisotropic unit cell. In order to deduce these 
results, the plaquette operator procedure has been generalised to a block 
operator technique which uses blocks higher than an unit cell and contains 
$f$-operator contributions acting only on a single central site of the block.
\end{abstract}

\section{Introduction}

The periodic Anderson model (PAM) is one of the basic models largely used in 
the study of strongly correlated systems whose properties can be 
described at the level of two effective bands,
like heavy-fermion systems \cite{int2}, intermediate-valence 
compounds \cite{int3}, or even high critical temperature superconductors 
\cite{int4}. The model contains a free $d$ band hybridized with a correlated
system of $f$ electrons for which the one-site Coulomb repulsion in the form 
of the Hubbard interaction is locally present. Seen from the theoretical side,
PAM has the peculiarity that even
its one dimensional Hamiltonian is sufficiently complicated to not allow the 
knowledge of its exact solutions even in 1D. As a consequence, taking into 
account that the exact description possibilities increase in difficulty with 
the increase of the dimensionality of the system in the physical region 
$D=1-3$, the physics provided by PAM is almost exclusively interpreted based 
on approximations. This situation enhance the difficulty of a good quality
theoretical analysis, since exact bench-marks in testing the approximations or
numerical simulations are almost completely missing. Because of this fact,
even the starting point of the theoretical description, the knowledge of the
ground-state is poorly developing. Given by this, efforts have been made
for the construction of exact ground-states at least in restricted regions
of the parameter space. In this frame, based on the observation that the 
infinitely repulsive case in relative terms is easier to treat, the first 
exact ground-states 
have been deduced at $U=\infty$, in restricted regions of the parameter space,
based on a work of several years \cite{exa1,exa2,exa3,exa4}. 

The first exact ground-states at finite value of the interaction have been 
published recently in 1D \cite{exa11,exa12}, and 2D \cite{exa13,exa13a}, 
respectively. These ground-states emerge on continuous but restricted 
regions of the $T=0$ phase diagram of the system which extend from the low 
$U$ limit up to the high $U$ limit as well. These solutions have been 
obtained by a 
decomposition of the Hamiltonian in positive semidefinite operators (PSO) as 
follows (i) the interaction term has been transformed into a PSO requiring at 
least one $f$ electron on every lattice site \cite{exa11}, and (ii) the 
remaining parts of the interaction term together with the one-particle 
components of the Hamiltonian have been transformed in PSO based 
on cell operators, using bonds in 1D \cite{exa11,exa12} or 
elementary plaquette operators in 2D \cite{exa13,exa13a,exa13b}. Two type 
of ground-states have been obtained in this manner: localized and itinerant 
once. The itinerant solution was found to emerge for imaginary hybridization 
matrix elements ($V_{\bf r}$), while the localized solution for real 
$V_{\bf r}$. Besides, the itinerant solution has been obtained only in the 
presence of anisotropic or distorted unit cells. The cell operators used in 
the transformation of the Hamiltonian had always the extension of an unit cell
(elementary plaquette in 2D), the $f$-creation operators being ,,uniformly'' 
considered, acting on all lattice sites of the elementary plaquette. 

We must note, that all exact ground-states deduced up today for PAM at 
finite value of the interaction $U$, based on the cell operators mentioned 
above, require the presence of the direct hopping $t^f \ne 0$ as well in the 
correlated $f$ band of the Hamiltonian, since the products of these cell
operators leading to PSO generate always $t^f \ne 0$ contributions. The 
presence of the direct $f$ hopping can be argued based on experimental data in
the case of $Pu$ \cite{Pu}, or some heavy-fermion compounds \cite{bev8}, but 
it remains in fact an extension term to the generic PAM Hamiltonian which does
not contain such type of contributions. Particularly for 2D, since for the 
$t^f = 0$ limit the results deduced in Refs.[\cite{exa13,exa13a}] are no more
valid, the case of the generic PAM Hamiltonian treated in exact terms remains
still a completely open problem.
 
Driven by the challenge to obtain exact solutions for the generic PAM in 2D,
the first questions which have to be clarified in these conditions are the 
following: Can we consider the ground-states deduced for PAM at finite $U$ in 
the presence of direct $f$ hopping in the Hamiltonian also potential 
ground-states for the generic PAM at finite $U$ (and $t^f = 0$) ? The unit cell
distortions and imaginary hybridization matrix elements are essential for the
emergence of itinerant phases ? This questions are important, since are 
connected to main problems related to PAM, for example the 
localized vs. itinerant behaviour of $f$-electrons, and of particles in general
in the system \cite{bev2,bev3,bev4,bev5,bev6,bev7}, or the interpretation of 
the PAM behaviour based exclusively on the local $f$-moment and its 
compensation (analyses made based on Kondo physics).  

Starting from this background, in this paper we are reporting the first exact 
ground-states for generic PAM in 2D at finite value of the interaction. As a
consequence, the deduction is made without the presence of the direct hopping 
terms for $f$ electrons in the Hamiltonian of the model ($t^f = 0)$. 
The obtained exact ground-state is 
itinerant, emerges for real values of the Hamiltonian parameters (including 
the hybridization matrix elements as well), is not necessarily connected to
distorted unit cells and presents clear similarities in its 
physical properties with the itinerant solution obtained in the presence of 
direct $f$ hopping terms in $\hat H$ \cite{exa13}. Since the momentum 
distribution function is continuous without any non-regularities in its 
derivatives of any order, the presented ground-state is a non-Fermi liquid 
state in the normal phase, possesses a large spin degeneracy, and globally
is paramagnetic. 

In order to deduce the reported results, major developments have been 
applied in comparison with the plaquette operator method used in 
Ref.(\cite{exa13}) and described in detail in Ref.(\cite{exa13a}), which has
been generalised to a block operator procedure. The introduced block is 
qualitatively different from the elementary plaquette previously used since
a) it has an extension greater than an unit cell and b) it contains $f$ 
creation operators acting only on one unique central site of the block. 
This choice allows us to represent the PAM Hamiltonian in term of PSO block 
products even in the absence of the direct $f$ hopping terms and to obtain 
non-localized ground-states in the presence of real hybridization matrix 
elements and absence of lattice distortions. All these are not possible to 
obtain in the frame of Refs.[\cite{exa13,exa13a}]. 

The consequences of the presented results are multiple. 
(i) At the level of exact solutions in 2D, the difference between the case 
$t^f \ne 0$ and $t^f=0$ in $\hat H$, (at least at the level of known exact 
ground-states) seems to not be extremely significant. Physically this can be 
understood based on the hybridization term which allows the movement of 
$f$-electrons even if direct hopping in the correlated band is not present in 
$\hat H$, and, in fact for the behaviour of the system (and also the 
ground-state), not the bare bands are important but the diagonalized once.  
For the presented case the $t^f\ne 0$ value seems only to shift the 
position in the parameter space of the emerging phase in comparison with the 
$t^f = 0$ situation, without changing essentially its physical properties. 
(ii) The fact that similar phases are obtained in exact terms for $t^f \ne 0$ 
and $t^f = 0$ as well shows that PAM can provide a behaviour different from 
Kondo physics. The statement is underlined as well by the absence of 
the exponential contributions characteristic for a Kondo type of behaviour in 
the exactly deduced ground-state energy values (see for example Ref. 
\cite{int2}). This possibility in fact exceeds the vicinity of a 
metal-insulator transition mentioned from this point of view in 
Ref.(\cite{exa13a}) since the presented phase diagram region extends 
continuously from the low $U$ limit to the high $U$ limit as well, up to 
$U\to \infty$.
(iii) Combining the here obtained result with the results previously deduced,
we observe that the delocalization of the electrons in PAM can emerge
simply because in this manner the system decreases its ground-state energy.
There are situations (depending on the parameters of the starting
Hamiltonian)\cite{exa12,exa13,exa13a,exa13b} when long-range density-density 
correlations are developing creating a localized phase, maintaining localized
as well the $f$ electrons within the system. When the delocalization occurs,
the long-range density-density correlation is lost.
(iv) We learn that using different type of blocks in constructing block 
operators in the process of the decomposition of the Hamiltonian into products
of positive semidefinite terms, different classes of Hamiltonians can be 
analysed. The block form, its extension, the uniform or non-uniform nature of 
the operator action inside the block, the type of dependence of the numerical 
coefficients entering in the block operator on the physical parameters (like 
spin), all are important in this description process. Based on this degree of
freedom, the method can be applied as well in the absence of direct $f$-hopping
terms as well.
(v) It becomes clear that in 
conditions in which the Hamiltonian remains hermitic, the real or imaginary 
(or even complex) hybridization matrix elements could provide similar physical
consequences. The nature of these matrix elements is given by the symmetry 
\cite{hybr1}, and in fact imaginary hybridization matrix elements has been 
used previously as well (see for example \cite{bev3}). 
(vi) The non-Fermi liquid properties present in rigorous terms in
the PAM phase diagram, at least in 2D are not essentially connected to 
imaginary hybridization matrix elements, nor direct $f$ hopping in the 
Hamiltonian, nor the presence of distorted unit cells. 

The remaining part of the paper has been constructed as follows. Section II.
presents the model and its description with block operators, Sect.III. 
describes the deduced exact ground-state, Sect.IV presents the conclusions, and
Appendix A. containing mathematical details closes the presentation.

\section{The presentation of the model.}

\subsection{The expression of the Hamiltonian.}

We start with a generic PAM Hamiltonian taken for a 2D lattice in the form
\begin{eqnarray}
\hat H = \sum_{k,\sigma} \epsilon^d_{{\bf k},\sigma} \hat d^{\dagger}_{{\bf k},
\sigma} \hat d_{{\bf k},\sigma} + E_f \sum_{{\bf k},\sigma} \hat f_{{\bf k},
\sigma} \hat f_{{\bf k},\sigma} + \sum_{{\bf k},\sigma} (V_{\bf k} 
\hat d^{\dagger}_{{\bf k},\sigma} \hat f_{{\bf k},\sigma} + H.c.) + \hat U \: ,
\label{h1}
\end{eqnarray}
where, the first term gives the kinetic energy for $d$-electrons, the
second term is the on-site $f$-electron energy, the third term represents the
hybridization, and the interaction term $\hat U$ describes the on-site Hubbard 
interaction written for $f$ electrons $\hat U = U \hat U_f$, 
$\hat U_f = \sum_{\bf i} \hat n^f_{{\bf i},\uparrow} \hat n^f_{{\bf i},
\downarrow}$, $U > 0$ being considered during this paper. As it can be seen, 
direct $f$-band is not present in the starting Hamiltonian.

For technical reasons, we transcribe $\hat H$ in ${\bf r}$ space, using
for the operators $\hat g = \hat d,\hat f$ the Fourier sum  $\hat g_{{\bf i},
\sigma} = \sum_{{\bf k}} e^{- i {\bf k} {\bf r}_{\bf i}} \hat g_{{\bf k},
\sigma}$. For this, we take into consideration at the level of the 
hybridization term, local ($V_0$) and non-local ($V_1$) nearest-neighbour 
contributions as well. The $d$-electron dispersion is taken into account 
including contributions up to next nearest-neighbour hoppings, which is not
unusual in the case of the study of real materials \cite{bev9}. The kinetic 
energy of $d$-electrons becomes
\begin{eqnarray}
\hat T_d &=& \sum_{{\bf i},\sigma}[t_x \hat d^{\dagger}_{{\bf i},\sigma}
\hat d_{{\bf i}+{\bf x},\sigma} + t_y \hat d^{\dagger}_{{\bf i},\sigma}
\hat d_{{\bf i}+{\bf y},\sigma} + t_{x+y} \hat d^{\dagger}_{{\bf i},\sigma}
\hat d_{{\bf i}+{\bf x}+{\bf y},\sigma} + t_{y-x} \hat d^{\dagger}_{{\bf i},
\sigma} \hat d_{{\bf i}+{\bf y}-{\bf x},\sigma} 
\nonumber\\
&+& t_{2x} \hat d^{\dagger}_{
{\bf i},\sigma} \hat d_{{\bf i}+2{\bf x},\sigma} + t_{2y} \hat d^{\dagger}_{
{\bf i},\sigma} \hat d_{{\bf i}+2{\bf y},\sigma} + H.c] \: ,
\label{h2}
\end{eqnarray}
where ${\bf x},{\bf y}$ represent the versors of the unit cell (see Fig.1), 
and the dispersion relation for $d$ electrons presented in Eq.(\ref{h1}) 
becomes
\begin{eqnarray}
\epsilon^d_{\bf k} = t_x e^{-i{\bf k}{\bf x}} + t_y e^{-i{\bf k}{\bf y}} +
t_{2x} e^{-2i{\bf k}{\bf x}} + t_{2y} e^{-2i{\bf k}{\bf y}} + 
t_{x+y} e^{-i{\bf k}({\bf x}+{\bf y})} + t_{y-x} e^{-i{\bf k}({\bf y}-{\bf x})
} + c.c.
\label{h3}
\end{eqnarray}
For the hybridization term we consider
\begin{eqnarray}
\hat V = \sum_{{\bf i},\sigma} [ V_0 \hat d^{\dagger}_{{\bf i},\sigma}
\hat f_{{\bf i},\sigma} + V_x (\hat d^{\dagger}_{{\bf i},\sigma} \hat f_{
{\bf i}+{\bf x},\sigma} + \hat f^{\dagger}_{{\bf i},\sigma} \hat d_{
{\bf i}+{\bf x},\sigma}) + V_y (\hat d^{\dagger}_{{\bf i},\sigma} \hat f_{
{\bf i}+{\bf y},\sigma} + \hat f^{\dagger}_{{\bf i},\sigma} \hat d_{
{\bf i}+{\bf y},\sigma}) + H.c. ] \: ,
\label{h4}
\end{eqnarray}
from where, the $V_{\bf k}$ hybridization matrix element from Eq.(\ref{h1})
can be expressed as
\begin{eqnarray}
V_{\bf k} = V_0 + (V_x e^{-i{\bf k} {\bf x}} + V_y e^{-i{\bf k} {\bf y}} +
c.c.) = V_0 + V_{1,{\bf k}} \: .
\label{h5}
\end{eqnarray}
Denoting by $\hat E_f = E_f \sum_{{\bf i},\sigma} \hat n^f_{{\bf i},\sigma}$ 
the on-site $f$-electron energy and using Eqs.(\ref{h2}-\ref{h5}), for the 
starting Hamiltonian presented in Eq.(\ref{h1}) we find
\begin{eqnarray}
\hat H = \hat T_d + \hat E_f + \hat V + \hat U \: ,
\label{h6}
\end{eqnarray}

The interaction term during this paper is exactly transformed in the form
\cite{exa11}
\begin{eqnarray}
\hat U_f = \sum_{\bf i} \hat n^f_{{\bf i},\uparrow} \hat n^f_{{\bf i},
\downarrow} = \hat P' + \sum_{\bf i} ( \sum_{\sigma} \hat n^f_{{\bf i},\sigma}
 - 1 ) \: ,
\label{h7}
\end{eqnarray} 
where, the positive semidefinite operator 
$\hat P' = \sum_{\bf i}( 1 - \hat n^f_{{\bf i},\uparrow} - \hat n^f_{{\bf i},
\downarrow} + \hat n^f_{{\bf i},\uparrow} \hat n^f_{{\bf i},\downarrow})$ 
defined by Eq.(\ref{h7}) requires for its lowest zero eigenvalue at least one 
$f$-electron on every lattice site. As will be clarified further on, the 
representation presented in Eq.(\ref{h7}) is a key feature from the point of 
view of the interaction term in the deduction of exact ground-states at
$U > 0$ presented here.

\subsection{The Hamiltonian written in term of block operators.}

Let us introduce connected to every site $i$ of the lattice the block
$A_{\bf i}$ as presented in Fig.1. It contains the lattice sites $i-1, i, i+1,
j, l$, being centred on the site $i$. The numbering of the sites inside the 
block is block independent and
given by the numbers $1,2,3,4,5$ in Fig. 1. The site-numbering inside the 
block is considered block-independent given by the translational symmetry 
of the system. Starting from the block $A_{\bf i}$ we are introducing block
operators as a sum of fermionic operators acting on the sites contained in 
$A_{\bf i}$ as follows
\begin{eqnarray}
\hat A_{{\bf i},\sigma} = a_{1,d} \hat d_{{\bf i},\sigma} + a_{2,d}
\hat d_{{\bf j},\sigma} + a_{3,d} \hat d_{{\bf i+1},\sigma} + a_{4,d}
\hat d_{{\bf l},\sigma} + a_{5,d} \hat d_{{\bf i-1},\sigma} + a_{1,f}
\hat f_{{\bf i},\sigma} \: ,
\label{h8}
\end{eqnarray}
where the $a_{n,d}$ and $a_{1,f}$ prefactors are numerical coefficients. 
Taking a block operator centred on another site $\hat A_{{\bf i}',\sigma}$, 
the indices of the fermionic operators follow the indices of the new sites
$(i'-1,i',i'+1,j',l')$, but for the numerical prefactors of $\hat A_{{\bf i}',
\sigma}$ we are keeping the same strategy in notation: $1$ represents the 
centre of the block, the index $2$ is of the site at the bottom of the block, 
the notation inside the block continuing anti clock-wise at the border of the 
block from $3$ up to the index $5$. 

The block operator $\hat A_{{\bf i},\sigma}$ used in the present description
has significant differences in comparison to the plaquette operators used 
previously \cite{exa13,exa13a}. Its novelty is twofold: (i) the here 
introduced block operator contains $f$-fermionic operators only in its unique 
central site, so contains a non-homogeneous f-operator action inside the block,
and (ii) the number of lattice sites per block being $5/4$, the used block has
an extension greater than an unit cell. These are key features which allow 
the study of the PAM Hamiltonian without the presence of the direct hopping 
terms at the level of $f$-electrons.  

Summing up now $\hat A^{\dagger}_{{\bf i},\sigma} \hat A_{{\bf i},\sigma}$ over
all lattice sites and taking periodic boundary conditions into account in 
both directions (for the result see Appendix A.), the following relation is 
obtained
\begin{eqnarray}
-\sum_{{\bf i},\sigma} \hat A^{\dagger}_{{\bf i},\sigma} 
\hat A_{{\bf i},\sigma} = \hat T_d + \hat V - \sum_{{\bf i},\sigma}
[|a_{1,f}|^2 \hat n^f_{{\bf i},\sigma} + (\sum_{n=1}^5 |a_{n,d}|^2) \hat n^d_{
{\bf i},\sigma} ] \: ,
\label{h9}
\end{eqnarray}
if the following equalities are present between the parameters of $\hat H$ and
the numerical prefactors of the block operators $\hat A_{{\bf i},\sigma}$
\begin{eqnarray}
&&- t_x = a^{*}_{1,d} a_{3,d} + a^{*}_{5,d} a_{1,d} \: , \quad
- t_y = a^{*}_{1,d} a_{4,d} + a^{*}_{2,d} a_{1,d} \: ,
\nonumber\\
&&- t_{x+y} = a^{*}_{5,d} a_{4,d} + a^{*}_{2,d} a_{3,d} \: , \quad
- t_{y-x} = a^{*}_{2,d} a_{5,d} + a^{*}_{3,d} a_{4,d} \: ,
\nonumber\\
&&- t_{2x} = a^{*}_{5,d} a_{3,d} \: , \quad
- t_{2y} = a^{*}_{2,d} a_{4,d} \: , \quad
- V_0 = a^{*}_{1,d} a_{1,f} \: , 
\nonumber\\
&&- V_x =a^{*}_{1,f} a_{3,d} = a^{*}_{5,d} a_{1,f} \: , \quad
- V_y =a^{*}_{1,f} a_{4,d} = a^{*}_{2,d} a_{1,f} \: .
\label{h10}
\end{eqnarray}
Taking into account that $\hat A_{{\bf i},\sigma} \hat A^{\dagger}_{{\bf i},
\sigma} + \hat A^{\dagger}_{{\bf i},\sigma} \hat A_{{\bf i},\sigma} =
|a_{1,f}|^2 + \sum_{n=1}^5 |a_{n,d}|^2$ and $\hat U = U \hat P' - U N_{\Lambda}
+ U \sum_{{\bf i},\sigma} \hat n^f_{{\bf i},\sigma}$, where $N_{\Lambda}$
represents the number of lattice sites, the Hamiltonian from Eq.(\ref{h6})
becomes
\begin{eqnarray}
\hat H = \sum_{{\bf i},\sigma} \hat A_{{\bf i},\sigma} \hat A^{\dagger}_{
{\bf i},\sigma} + U \hat P' + \hat R \: ,
\label{h11}
\end{eqnarray}
where $\hat R = - U N_{\Lambda} - 2 N_{\Lambda} (|a_{1,f}|^2 + \sum_{n=1}^5
|a_{n,d}|^2) + K \hat N$. In this expression $\hat N$ is the operator of the 
total number of electrons, and the constant $K$ is given by
\begin{eqnarray}
E_f + U + |a_{1,f}|^2 = \sum_{n=1}^5 |a_{n,d}|^2 = K \: .
\label{h12}
\end{eqnarray}
The Hamiltonian contained in Eq.(\ref{h11}) will be analysed in detail below.
We mention that the transformation of the starting $\hat H$ from Eq.(\ref{h6})
to the studied $\hat H$ from Eq.(\ref{h11}) is possible only if the Hamiltonian
parameters (considered known variables) satisfy the equations Eqs.(\ref{h10},
\ref{h12}). On their turn, Eqs.(\ref{h10},\ref{h12}) determine the unknown 
block operator parameters $a_{1,f},a_{1,d},a_{2,d},...,a_{5,d}$ in term of
Hamilton operator parameters that must be considered known variables. This 
works only for the case 
in which $\hat H$ from Eq.(\ref{h6}) can be written in the form of $\hat H$
presented in Eq.(\ref{h11}). Since the number of unknown variables is less than
the number of Hamiltonian parameters, solutions for the block operator 
parameters will be present only if some inter-dependences between $t_{\bf r},
V_{\bf r},E_f,U$ are present. These inter-dependences fix the parameter space
region ${\cal{P}_H}$ in which the here obtained results are valid (see 
Fig. 2. and the explications from Sect.III.B.).

\section{Exact ground-state wave-function solution.}
 
\subsection{The derivation of the exact ground-state.}

We are studying the Hamiltonian from Eq.(\ref{h11}) at a fixed number of
particles $N$ in the system. As a consequence, since $N$ is a constant of
motion, $\hat H$ becomes $\hat H = \hat P + E_g$, where the positive 
semidefinite operator $\hat P = \sum_{{\bf i},\sigma} \hat A_{{\bf i},\sigma}
\hat A^{\dagger}_{{\bf i},\sigma} + U \hat P'$ has zero minimum eigenvalue, and
the constant $E_g$ is given by $E_g = K N - N_{\Lambda}(U + 2|a_{1,f}|^2 +
2\sum_{n=1}^5 |a_{n,d}|^2)$. In these conditions, the ground-state wave 
function of the model inside ${\cal{P_H}}$ is defined via
$\hat P |\Psi_g\rangle = 0$. To find $|\Psi_g\rangle$, we have to keep in mind
that $\hat P'$ requires for its minimum (and zero) eigenvalue at least one 
$f$-electron on every lattice site, and that the introduced block operators
satisfies the following properties
\begin{eqnarray}
\hat A_{{\bf i},\sigma}^{\dagger} \hat A_{{\bf i},\sigma}^{\dagger} = 0 \: ,
\quad \hat A_{{\bf i},\sigma}^{\dagger} \hat A_{{\bf j},\sigma'}^{\dagger} = 
- \hat A_{{\bf j},\sigma'}^{\dagger} \hat A_{{\bf i},\sigma}^{\dagger} \: .
\label{h13}
\end{eqnarray}
Starting from Eq.(\ref{h13}) we observe that the block operator part of 
Eq.(\ref{h11}) applied to $\prod_{\bf i} \hat A^{\dagger}_{{\bf i},\uparrow}
\hat A^{\dagger}_{{\bf i},\downarrow}$ gives zero. Furthermore, given by the
presence of $\hat P'$ in $\hat H$, we add to the ground-state 
the contribution $ \hat F_{\mu} = \prod_{\bf i} 
(\mu_{{\bf i},\uparrow} \hat f^{\dagger}_{{\bf i}, \uparrow} + 
\mu_{{\bf i},\downarrow} \hat f^{\dagger}_{{\bf i}, \downarrow})$, where 
$\mu_{{\bf i},\sigma}$ are arbitrary coefficients. As a consequence, the 
ground-state wave-function with the property $\hat P |\Psi_g\rangle = 0$ 
becomes
\begin{eqnarray}
|\Psi_g \rangle = \prod_{\bf i} [\hat A^{\dagger}_{{\bf i},\uparrow} 
\hat A^{\dagger}_{{\bf i},\downarrow}(\mu_{{\bf i},\uparrow} \hat f^{
\dagger}_{{\bf i}, \uparrow} + \mu_{{\bf i},\downarrow} \hat f^{\dagger}_{
{\bf i}, \downarrow})] |0 \rangle \: ,
\label{h14}
\end{eqnarray}
where, $|0\rangle$ is the bare vacuum with no fermions present. The 
wave-function presented in Eq.(\ref{h14}) is the first exact ground-state
obtained for generic PAM at finite $U$. The product in Eq.(\ref{h14}) must be 
taken over all lattice sites. 
Because of this reason, the product of the creation 
operators in Eq.(\ref{h14}) introduces $N = 3 N_{\Lambda}$ particles into the 
system, so $|\Psi_g\rangle$ corresponds to $3/4$ filling. All degeneration 
possibilities of the ground-state are contained in Eq.(\ref{h14}), since the 
wave function with the property $\hat P |\Psi\rangle = 0$ at $3/4$ filling 
always can be written in the presented $|\Psi_g\rangle$ form. We however 
underline that PAM contains two hybridized bands, and $3/4$ filling for a 
two-band system means in fact half filled upper hybridized band (the lower 
band being completely filled up). We mention, that $|\Psi_g\rangle$ 
describes rigorously only the $U > 0$ case, since the presence of the 
$\hat F_{\mu}$ operator into the ground-state is just required by the 
non-zero $U$ value. As a consequence, the ground-state at $U=0$ cannot be 
expressed in the form presented in Eq.(\ref{h14}).

\subsection{Solutions for the block operator parameters.}

We are now interested to find the $T=0$ phase diagram region where the 
solution from Eq.(\ref{h14}) is valid. For this reason the system of equations
Eqs.(\ref{h10},\ref{h12}) must be solved for the block operator parameters.
Solving the problem, we are considering all $\hat H$ parameters real.
From the $V_x,V_y$ components of Eq.(\ref{h10}) we find $a^{*}_{5,d} = a_{3,d}
(a^{*}_{1,d}/a_{1,d}), \: a^{*}_{4,d} = a_{2d} (a^{*}_{1,d}/a_{1,d})$, and 
introducing the anisotropy parameter $\chi = a_{2,d}/a_{3,d} = t_y/t_x$
(which must be real since $t_y/t_x$ is real), we 
realize that all $y$ components of $\hat H$ parameters can be expressed via 
$x$ components and $\chi$ as follows
\begin{eqnarray}
t_y = \chi t_x \: , \quad V_y = \chi V_x \: , \quad, t_{2y} = \chi^2 t_{2x}
\: , \quad |t_{x+y}| = 2 |\chi t_{2x}| \: .
\label{h15}
\end{eqnarray}
Solutions are obtained for $|t_{x+y}| = |t_{y-x}|$, $sign(\chi) = - 
sign(t_{x+y})$, $sign(t_x) = sign(V_0) sign(V_x)$, and the remaining equations
for the $x$ components of the $\hat H$ parameters provide the solutions
\begin{eqnarray}
|a_{1,d}| = \frac{|t_x|}{2\sqrt{|t_{2x}|}} \: , |a_{2,d}| = |a_{4,d}| =
|\chi| |a_{3,d}| = |\chi| |a_{5,d}| = |\chi|\sqrt{|t_{2x}|} \: ,
\quad |a_{1,f}| = 2 \sqrt{|t_{2x}|} \frac{|V_0|}{|t_x|} \: ,
\label{h16}
\end{eqnarray}
all $a_{n,d},a_{1,f}$ being considered real.
Introducing the notations $t=|t_{2x}/t_{x}|$, and $v=|V_0/t_x|$, the solutions
require $|V_0/V_x|=1/(2t)$ and are situated in the parameter space on the 
surface
\begin{eqnarray}
\frac{E_f+U}{|t_x|} = \frac{1}{4t} + 2 t [1 + \chi^2 - v^2] \: .
\label{h17}
\end{eqnarray}
This surface is presented (for $\chi=1.5$) in Fig. 2. As can be seen, it 
extends from the small $U$ domain continuously to the high $U$ domain up to
$U\to \infty$ in the $t\to 0$ limit. Modifying $\chi$, the general shape of
the obtained phase diagram region will not be changed. To be situated inside 
the phase diagram region ${\cal{P_H}}$ where the reported 
ground-state occurs for example in the isotropic case $\chi=1$, (which means
$t_1 = t_x = t_y, \: \: t_2 = t_{2x} = t_{2y}, \: \: t_2' = t_{x+y} = t_{y-x},
\: \: V_1 = V_x = V_y$), the parameters $t_1,t_2,V_0,U$ can be arbitrarily 
chosen, and $|t_2'| = 2 |t_2|, \: \: |V_1|=2 |V_0 t_2/t_1|$ must hold together
with Eq.(\ref{h17}) which determine $E_f/|t_1|$. As can be observed, 
${\cal{P_H}}$ can be reached by quite physical $\hat H$ parameter values.

\subsection{Physical properties of the obtained solutions.}

The magnetic properties of the wave-function of the form presented in 
Eq.(\ref{h14}) have been analysed in detail previously \cite{exa13,exa13a}.
Here the expression of $\hat A_{{\bf i},\sigma}$ is completely new, but the
described techniques can be well applied. 
Given by the arbitrary nature of the $\mu_{{\bf i},\sigma}$ coefficients, 
$|\Psi_g\rangle$ possesses a large spin degeneracy in the total spin of the 
system, being globally paramagnetic.

Studying the particle number distribution on different sites created by the
$\prod_{\bf i}$ product over the creation operators in $|\Psi_g\rangle$ from 
Eq.(\ref{h14}) together with the
concrete block operator presented in Eq.(\ref{h8}), it turns out that the
obtained ground-state wave-function contains different contributions with one,
two and three particles per site in the lattice. As a consequence, the system 
described by
$|\Psi_g\rangle$ is not characterised by an uniform particle distribution,
the expectation value of the hopping matrix elements and non-local 
hybridizations is non-zero, so the system is not localized and the electrons 
in the ground-state are itinerant. All these information show that the deduced
ground-state is an itinerant paramagnet.

The deduced ground-state being itinerant, its properties can be easier 
described using a ${\bf k}$-space representation. Starting this, for the  
Fourier transform of the block operators we find
\begin{eqnarray}
\hat A^{\dagger}_{{\bf i},\sigma} = \sum_{\bf k} e^{i{\bf k} {\bf r}_{\bf i}}
(a^{*}_{{\bf k},d} \hat d^{\dagger}_{{\bf k},\sigma} + a^{*}_{1,f} \hat f^{
\dagger}_{{\bf k},\sigma}) \: ,
\label{a1}
\end{eqnarray}
where $a^{*}_{{\bf k},d} = a^{*}_{1,d} + a^{*}_{2,d} e^{-i{\bf k}{\bf y}} +
a^{*}_{3,d} e^{+i{\bf k}{\bf x}} + a^{*}_{4,d} e^{+i{\bf k}{\bf y}} + a^{*}_{
5,d} e^{-i{\bf k}{\bf x}}$. Using now the definition of $\epsilon^d_{\bf k}$
and $V_{\bf k}$ from Eqs.(\ref{h3},\ref{h5}), we observe that exactly when the
presented solution holds (i.e. Eqs.(\ref{h10},\ref{h12}) are satisfied), we
obtain
\begin{eqnarray}
V_{\bf k} = - a_{1,f} a^{*}_{{\bf k},d} \: , \quad \epsilon^d_{\bf k} = 
K - |a_{{\bf k},d}|^2 \: .
\label{a2}
\end{eqnarray}
Using the notation $\Delta_{\bf k} = \sqrt{|a_{1,f}|^2 + |a_{{\bf k},
d}|^2}$, we introduce new canonical Fermi operators
\begin{eqnarray}
\hat C^{\dagger}_{1,{\bf k},\sigma} = \frac{1}{\Delta_{\bf k}} (a_{1,f}^{*}
\hat d^{\dagger}_{{\bf k},\sigma} - a^{*}_{{\bf k},d} \hat f^{\dagger}_{
{\bf k},\sigma}) \: , \quad
\hat C^{\dagger}_{2,{\bf k},\sigma} = \frac{1}{\Delta_{\bf k}} (a_{{\bf k},
d}^{*} \hat d^{\dagger}_{{\bf k},\sigma} + a^{*}_{1,f} \hat f^{\dagger}_{
{\bf k},\sigma}) \: ,
\label{a3}
\end{eqnarray}
which satisfy standard Fermionic anti-commutation rules. Using now Eqs.(
\ref{a2},\ref{a3}), we realize that $\hat H$ from Eq.(\ref{h1}) becomes
\begin{eqnarray}
\hat H = \sum_{{\bf k},\sigma} E_{1,{\bf k}} \hat C^{\dagger}_{1,{\bf k},
\sigma} \hat C_{1,{\bf k},\sigma} + \sum_{{\bf k},\sigma} E_{2,{\bf k}}
\hat C^{\dagger}_{2,{\bf k},\sigma} \hat C_{2,{\bf k},\sigma} + U\hat P'
- U N_{\lambda} \: ,
\label{a4}
\end{eqnarray}
where $E_{1,{\bf k}} = K =$ constant, and $E_{2,{\bf k}} = - K + E_f + U + 
\epsilon^d_{\bf k}$, $E_{1,{\bf k}} - E_{2,{\bf k}} = \Delta_{\bf k} > 0$.
Since for the ground-state $\hat P' |\Psi_g\rangle = 0$, we obtain 
$\hat H_{gr}$, the Hamiltonian exactly diagonalized for the ground-state, 
in the form
\begin{eqnarray}
\hat H_{gr} = \sum_{{\bf k},\sigma} K \hat C^{\dagger}_{1,{\bf k},
\sigma} \hat C_{1,{\bf k},\sigma} + \sum_{{\bf k},\sigma} E_{2,{\bf k}}
\hat C^{\dagger}_{2,{\bf k},\sigma} \hat C_{2,{\bf k},\sigma} - U N_{\lambda} 
\: .
\label{a5}
\end{eqnarray}
So, for the deduced ground-state (i.e.in the parameter space inside 
${\cal{P_H}}$), the Hamiltonian can be mapped into a 
two-band Hamiltonian with separated bands (determined also by $U$) 
whose upper band is completely flat (note that the 
starting $d$-band in Eq.(\ref{h1}) is with dispersion and there is no hopping
present for $f$-electrons in the starting Hamiltonian). Because of 
$N = 3 N_{\Lambda}$, the upper flat band (in which fermions are created by
$\hat C^{\dagger}_{1,{\bf k},\sigma}$) is half filled, and the lower band
(in which $\hat C^{\dagger}_{2,{\bf k},\sigma}$ creates particles) is 
completely filled up. As a consequence, denoting the ground-state expectation 
values by $\langle ... \rangle = \langle \Psi_g|...|\Psi_g\rangle/\langle 
\Psi_g|\Psi_g \rangle$, we have (see also \cite{exa13})
\begin{eqnarray}
\langle \hat C^{\dagger}_{1,{\bf k},\sigma} \hat C_{1,{\bf k},\sigma} \rangle
= \frac{1}{2} \: , \quad
\langle \hat C^{\dagger}_{2,{\bf k},\sigma} \hat C_{2,{\bf k},\sigma} \rangle
= 1 \: .
\label{a6}
\end{eqnarray}
We also mention that since the lower band is completely filled up, $\langle
\hat C^{\dagger}_{2,{\bf k},\sigma} \hat C_{1,{\bf k},\sigma} \rangle = 0$.
The second relation from Eq.(\ref{a6}) is trivial from physical point of view
since the lower band is completely filled up. Contrary to this, the first 
relation (see Fig. 3.) shows a momentum distribution function for the upper 
band without non-regularities in its derivatives of any order, signalling a 
clear non-Fermi liquid type of behaviour in 2D deduced in exact terms. 
The deduced phase is present also in the isotropic case ($\chi =1$), and
we underline that the result has clear physical signification even in the 
case in which ${\cal{P_H}}$ behaves complete repulsively from RG point of 
view \cite{x1}.

The novelty of this result in comparison with the behaviour reported in 
\cite{exa13} is threefold:
(i) here we are situated in generic PAM (i.e. $t^f=0$);
(ii) the hybridization matrix element is real; and 
(iii) distorted unit cell is not necessary for the emergence of the itinerant 
phase. As a consequence, the presence of the deduced behaviour is much 
more general then suggested by our previous work. 

Expressing $\hat f_{{\bf k},\sigma}, \hat d_{{\bf k},\sigma}$ from 
Eq.(\ref{a3}), and using Eq.(\ref{a6}), all needed ground-state expectation 
values connected to $\hat H$ in term of the starting fermionic operators can 
be deduced. Introducing the notation $I({\bf k}) = |V_{\bf k}|^2/(|a_{1,f}|^4 +
|V_{\bf k}|^2)$, we find
\begin{eqnarray}
\langle \hat f^{\dagger}_{{\bf k},\sigma} \hat f_{{\bf k},\sigma} \rangle =
1-\frac{1}{2} I({\bf k}) \: , \quad
\langle \hat d^{\dagger}_{{\bf k},\sigma} \hat d_{{\bf k},\sigma} \rangle =
\frac{1}{2}+\frac{1}{2} I({\bf k}) \: , \quad 
V_{\bf k} \langle \hat d^{\dagger}_{{\bf k},\sigma} \hat f_{{\bf k},\sigma} 
\rangle = - \frac{1}{2} |a_{1,f}|^2 I({\bf k}) \: .
\label{a7}
\end{eqnarray}
Based on Eq.(\ref{a7}) it can be checked that $n^f_{\bf k}$ and $n^d_{\bf k}$
as well are free from non-regularities in the whole first Brillouin zone.
The ground-state energy becomes $E_g/N_{\Lambda} = - U -2 |a_{1,f}|^2 +
\sum_{n=1}^{5} |a_{n,d}|^2$.
 
\section{Summary and conclusions}

We are presenting for the first time exact ground-states for the generic 
periodic Anderson model (PAM) at finite on-site repulsion for $f$-electrons 
$U$, in $D=2$ dimensions, the Hamiltonian not containing 
direct hopping terms for $f$-electrons $(t^f = 0)$. For this reason, 
and on this line
a) we generalised the previously used elementary plaquette operators
\cite{exa13,exa13a} to a block operator containing non-uniform $f$-operator
contributions (the $f$-operators acting only on one site of the block);
b) the block itself has been chosen to have an extension higher than an unit 
cell, so it cannot be considered elementary plaquette operator as used in 
previous studies;
c) based on the presented developments it was possible to analyse for the 
first time for PAM the $t^f=0$ generic case at finite $U$ and 2D in exact 
terms (in restricted regions of the parameter space);
d) again based on points a),b) an itinerant state holding non-Fermi liquid 
properties and presenting similarities with the itinerant state deduced in
Ref.[\cite{exa13}] has been obtained without distortions in the system and
presence of real hybridization matrix elements (we underline that results of 
the type mentioned in the points c),d) cannot be deduced in the frame of Refs.
(\cite{exa13,exa13a}));
e) the deduced exact results allow to present the first exact phase diagram
region for the generic PAM at finite $U$ and 2D, which is not included in the
previously deduced phase diagram regions;
f) starting from the presented technique and results we learn that using
different type of blocks in constructing block operators, different classes of
Hamiltonians can be analysed. The block form, its extension, the uniform or
non-uniform nature of the operator action inside the block, the spin 
dependence or non-dependence of the numerical coefficients of the block
operator, all are important in this description process;
g) the fact that at least in some regions of the parameter space (which 
however extend from the low $U$ to the high $U$ limit), 
similar phases are obtained in exact terms at $t^f \ne 0$
and $t^f = 0$ for PAM, underlines that this model can provide a behaviour 
differently from Kondo physics (where only the local f-moments and its 
compensation play the main role). This can happen not only in the region of a 
potential metal-insulator transition but also elsewhere in the phase diagram.

\acknowledgements

The research has been supported by contract OTKA-T-037212 and FKFP-0471 of 
Hungarian founds for scientific research. The author kindly acknowledge 
extremely valuable discussions on the subject with Dieter Vollhardt.
He also would like to thank for the kind hospitality of the Department 
of Theoretical Physics III., University Augsburg in autumn 2001, 4 months of  
working period relating this field spent there, and supported by Alexander 
von Humboldt Foundation. 

\newpage
\appendix

\section{The plaquette operator contributions summed up over the lattice 
sites.}

The expression $\hat A^{\dagger}_{{\bf i},\sigma} \hat A_{{\bf i},\sigma}$ 
summed up over the whole lattice considered with periodic boundary conditions 
in both directions is presented below.
\begin{eqnarray}
&&\sum_{{\bf i},\sigma} \hat A^{\dagger}_{{\bf i},\sigma} \hat A_{{\bf i},
\sigma} =
\nonumber\\
&&\sum_{{\bf i},\sigma}  \{
[\hat d^{\dagger}_{{\bf i},\sigma} \hat d_{{\bf i} + {\bf x}, \sigma}
(a^{*}_{1,d} a_{3,d} + a^{*}_{5,d} a_{1,d})
+ H.c.] +
[\hat d^{\dagger}_{{\bf i},\sigma} \hat d_{{\bf i} + {\bf y}, \sigma}
(a^{*}_{1,d} a_{4,d} + a^{*}_{2,d} a_{1,d}) + H.c.] +
\nonumber\\
&&[\hat d^{\dagger}_{{\bf i},\sigma} \hat d_{{\bf i} + ({\bf x}+{\bf y}), 
\sigma} (a^{*}_{5,d} a_{4,d} + a^{*}_{2,d} a_{3,d}) + H.c.] +
[\hat d^{\dagger}_{{\bf i},\sigma} \hat d_{{\bf i} + ({\bf y}-{\bf x}), 
\sigma} (a^{*}_{2,d} a_{5,d} + a^{*}_{3,d} a_{4,d}) + H.c.] +
\nonumber\\
&&[\hat d^{\dagger}_{{\bf i},\sigma} \hat d_{{\bf i} + 2{\bf y},\sigma} 
a^{*}_{2,d} a_{4,d} +
\hat d^{\dagger}_{{\bf i},\sigma} \hat d_{{\bf i} + 2{\bf x}, \sigma} 
a^{*}_{5,d} a_{3,d} + H.c.] +
\nonumber\\
&&[\hat f^{\dagger}_{{\bf i},\sigma} \hat d_{{\bf i} + {\bf x},\sigma} 
a^{*}_{1,f} a_{3,d} +
\hat d^{\dagger}_{{\bf i},\sigma} \hat f_{{\bf i} + {\bf x}, \sigma} 
a^{*}_{5,d} a_{1,f} + H.c.] +
\nonumber\\
&&[\hat f^{\dagger}_{{\bf i},\sigma} \hat d_{{\bf i} + {\bf y},\sigma} 
a^{*}_{1,f} a_{4,d} +
\hat d^{\dagger}_{{\bf i},\sigma} \hat f_{{\bf i} + {\bf y}, \sigma} 
a^{*}_{2,d} a_{1,f} + H.c.] +
\nonumber\\
&&\hat f^{\dagger}_{{\bf i},\sigma} \hat f_{{\bf i}, \sigma} |a_{1,f}|^2 +
\hat d^{\dagger}_{{\bf i},\sigma} \hat d_{{\bf i}, \sigma} 
(\sum_{n=1}^5 |a^{*}_{n,d}|^2 ) +
[\hat d^{\dagger}_{{\bf i},\sigma} \hat f_{{\bf i},\sigma} 
a^{*}_{1,d} a_{1,f} + H.c.] \} \: .
\label{b1}
\end{eqnarray}

\newpage

\begin{figure}[h]
\centerline{\epsfbox{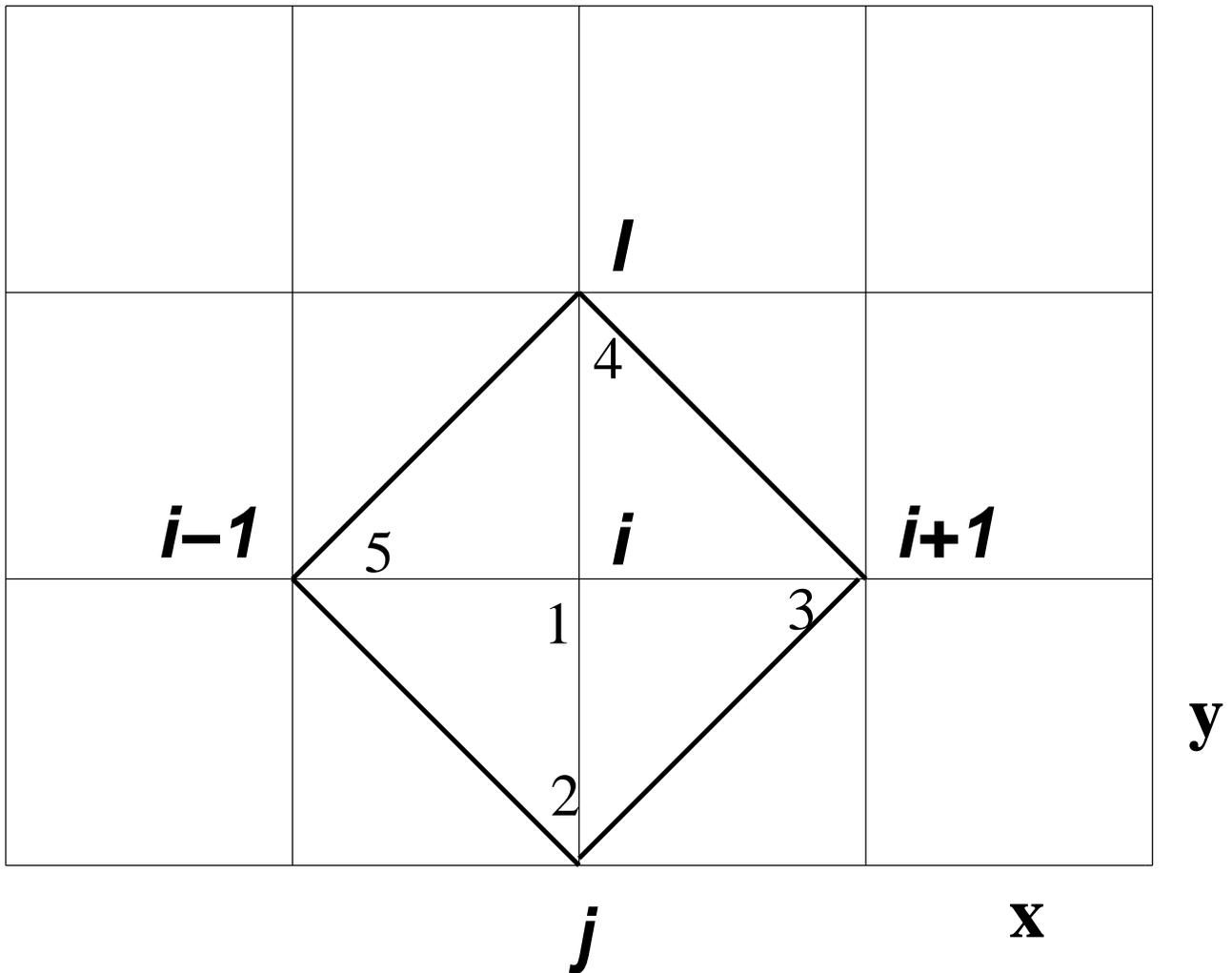}}
\caption{The block $A_{\bf i}$ (thick line) centred on the site ${\bf i}$ of 
the lattice. ${\bf x}$ and ${\bf y}$ are the versors of the unit cell.}
\label{fig1}
\end{figure}

\newpage

\begin{figure}[h]
\centerline{\epsfbox{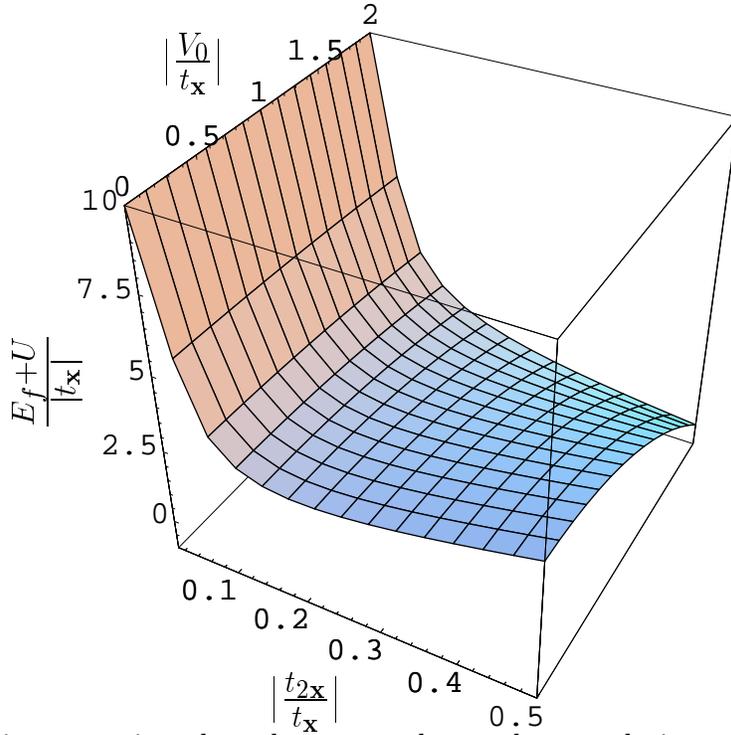}}
\caption{Phase diagram region where the presented ground-state solution occurs.
The anisotropy parameter is $\chi=1.5$ and the relation $|V_0/V_x|=1/(2|t|)$ 
with $t=t_{2{\bf x}}/t_{\bf x}$ must also hold. For $t\to 0$ the presented
surface extends along the $z=(E_f+U)/|t_{\bf x}|$ axis to $z\to \infty$.}
\label{fig2}
\end{figure}

\newpage

\begin{figure}[h]
\centerline{\epsfbox{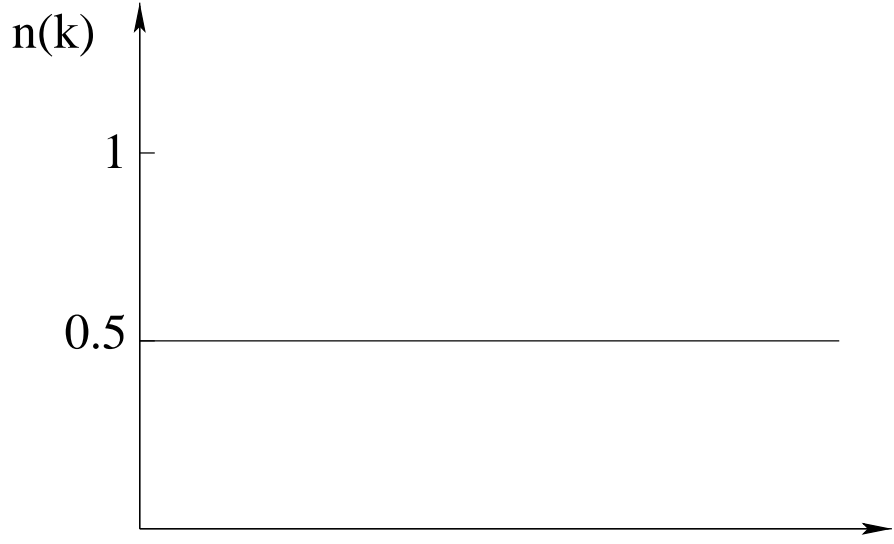}}
\caption{Momentum distribution function 
$n({\bf k}) = \langle \hat C^{\dagger}_{1,{\bf k},\sigma} \hat C_{1,{\bf k},
\sigma} \rangle$ for the upper half filled diagonalized band, taken over the
whole first Brillouin zone.}  
\label{fig3}
\end{figure}

\end{document}